\documentclass[aps,twocolumn,pra,floatfix]{revtex4}
\usepackage{epsfig}
\usepackage{graphicx}
\usepackage{dcolumn}
\usepackage{amssymb,amsmath}
\usepackage{mathrsfs}
\begin{document}

	
\title{Long-range parity non-conserving electron-nucleon interaction}
	
\author{V. A. Dzuba, V. V. Flambaum, P.  Munro-Laylim}

\affiliation{School of Physics, University of New South Wales, Sydney 2052, Australia}

\begin{abstract}
As known, electron vacuum polarization by nuclear Coulomb field produces Uehling potential with the range $\hbar/2m_e c$. Similarly, neutrino vacuum polarization  by $Z$ boson field produces long range potential $\sim G^2/r^5$  with the large range $\hbar/2m_{\nu}c$. Attempts to measure parity-conserving part of this potential produced only limits on this potential which are several orders of magnitude higher than the standard model predictions. We show that parity  non-conserving (PNC) part of the neutrino exchange potential $W_L(r)$  gives a significant fraction of the observed PNC effects. Mixed $Z-\gamma$ electron vacuum polarization  produces PNC potential with range $\hbar/2m_e c$, which exceeds the range of the weak interaction by five orders of magnitude. We calculate contribution  of the long-range PNC  potentials to the nuclear spin independent and nuclear spin dependent PNC effects. 
 The cases of the single-isotope PNC effects  and the ratio of PNC effects in different isotopes  are considered for Ca, Cs, Ba, Sm, Dy, Yb, Hg, Tl, Pb, Bi, Fr, Ra atoms and ions. Contributions of the long-range PNC potentials ($\sim$1\%) significantly exceed experimental error (0.35\%) for PNC effect in Cs.  
\end{abstract}

\maketitle
	
\section{Introduction}\label{SectionIntroduction}

\begin{figure}[tbh]
	\centering
	\includegraphics[scale=0.8]{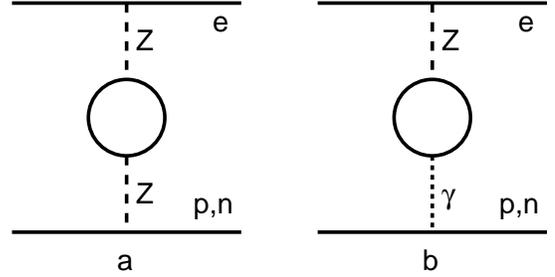}
	\caption{a. Vacuum polarization by the nuclear weak Z-boson field which produces long range parity violating potential $W_L(r)\propto 1/r^5$ . In the case of neutrino loop the range is $r_{\nu}=\hbar/(2 m_{\nu} c)$.
	b. Mixed $Z-\gamma$ vacuum polarization  which produces long range parity violating potential $W_L(r)\propto 1/r^3$. In the case of electron  loop the range is $r_e=\hbar/(2 m_{e} c)$.}
	\label{fig}
\end{figure}

As it was firstly noted by Feynman \cite{Feynman} and calculated in  Refs.~\cite{FeinbergPR1968,Feinberg1989,HsuPRD1994}, exchange by two neutrinos ( see e.g. diagram on Fig.~\ref{fig}~a) produces long range potential $\sim G^2/r^5$, where $G$ is Fermi constant. However, effects of parity conserving part of this potential are many orders of magnitude smaller than sensitivity of experiments Refs.~\cite{Kapner2007,Adelberger2007,Chen2016,Vasilakis2009,Terrano2015,StadnikPRL2018}.

In Ref. \cite{Ghosh} it was noted the neutrino exchange potential has parity non-conserving (PNC) part. Earlier it was demonstrated that  mixed $Z-\gamma$ electron vacuum polarization  produces PNC potential with the  range $\hbar/2m_e c$ (see Fig.~\ref{fig}~b) , which exceeds the range of the weak interaction by five orders of magnitude \cite{FS07}. In the present  paper we show that the contributions of the long-range PNC potentials to PNC effects in atoms is $\sim$1\% and this significantly exceeds the error 0.35\% of the PNC measurement in Cs atom ~\cite{Wood} and the  error $<$0.5\% in the many-body atomic  calculations of the $Z$-boson contribution ~\cite{DzuFlaSus89a,DzuFlaGin02,Blundell90,Blundell92,Ginges05,PBD09,PBD10,CsPNC12}. The work is in progress to improve both, experimental~\cite{Elliott1,Elliott2,Elliott3} and theoretical~\cite{TXD22} accuracy. 

The error in atomic calculations cancels out in  the ratio of the PNC amplitudes in different isotopes of the same atom~\cite{DzuFlaKhr86,BDF09,VF19}.
The work is in progress for such measurements too, in particular for the chain of isotopes of Yb atom~\cite{Budker19}.
The study of the parity non-conservation (PNC) in atoms play important role in testing the standard model (SM) and searching for new physics beyond it \cite{RevPNC,RevPNC1}.


Radiative corrections to the PNC amplitudes  of the order $\alpha \approx 1/137$ have been presented as the radiative corrections to proton and neutron weak charges and exceed 1\%  \cite{Marciano}. Weak charge itself is the constant of the electron-nucleon weak interaction due to the $Z$-boson exchange which has interaction range $r_Z=\hbar/M_Z c=0.002$ fm. On the nuclear and atomic scales this may be considered as a Fermi-type contact interaction. However,  radiative corrections actually generate PNC interaction of a  much longer range.  Neutrino vacuum polarization  by the nuclear weak $Z$ boson field (see Fig.~\ref{fig}~a) produces PNC potential $W_L(r)\propto 1/r^5$  which has exponential cut-off on the distance $r_{\nu}=\hbar/(2 m_{\nu} c)$ exceeding atomic size by many orders of magnitude. Mixed $Z-\gamma$ electron vacuum polarization (see Fig.~\ref{fig}~b)  induces PNC interaction $\propto 1/r^3$  of the range $r_e=\hbar/(2 m_e c)=193$ fm \cite{FS07}, similar to the range of the Uehling  potential due to electron vacuum polarization by the nuclear Coulomb field.

 The deviation from the contact limit for this long-range PNC interaction  may be significant since in heavy atoms relativistic  Dirac electron wave functions rapidly increase toward the nucleus ($\psi_{s1/2}\psi_{p1/2} \propto 1/r^{2-2 \gamma}$, 
 where $Z$ is the nuclear charge,$\gamma=\sqrt{1 - Z^2 \alpha^2}$, so $2-2 \gamma\approx Z^2 \alpha^2$).  
  This rapid variation of the electron wave function between $r_e=\hbar/(2 m_e c)$  and the nucleus requires proper treatment of the long-range PNC potential $W_L$.  Contrary to the contact PNC interaction $W_Q$, potential $W_L$ gives direct contribution  to the matrix elements between electron orbitals with angular momentum  higher than $l=0$ and $l=1$. Note that   in Yb the  PNC mixing between dominating configurations is given by the $\langle p|W|d \rangle$ matrix element.  Therefore, this qualitative feature of the long range PNC interaction also should be investigated. 
 
 Note that deviation of the contribution of the long range potential $W_L$ from its contact interaction limit is roughly proportional to $\alpha (Z \alpha)^2 $. Indeed, in the non-relativistic limit ($Z \alpha \ll 1$) an  $s$-wave function  and gradient of a $p$ -wave function are approximately  constant near the nucleus  and the PNC matrix element $\langle s|W_L| p \rangle$ is not sensitive to the range of the potential as soon it is much smaller than $a_B/Z$.   
Other contributions of the order  $\alpha (Z \alpha)$ may be found in paper \cite{Ginges05} and references therein.

Note that the electron-positron  loop may be replaced by the particle-hole pair corresponding to the excitation of electron from the atomic  core. However, this is a correction which has already been included in the many-body calculations of the PNC effects. 
A different mechanism of the long-range PNC interaction between an atom and charged particle (via PNC vector polarizability) has been discussed in Ref.  \cite{Flambaum92}.

In the present paper we consider corrections due to long range PNC interaction to the PNC amplitudes in many atoms of experimental interest. We consider the cases of single isotope measurements and the ratio of the PNC amplitudes for a chain of isotopes. We perform calculations of the nuclear spin independent (NSI) interaction and the nuclear spin dependent (NSD)  interaction. 

\section{Long-range PNC potential due to the mixed photon - $Z$ vacuum polarization}

It was suggested in Ref.~\cite{FS07} that  photon-Z-boson mixing via electron loop (see Fig.~\ref{fig}~b) leads to the long-range parity non-conserving potential.  In Ref.~\cite{FS07} this potential was obtained for a point-like nucleus and contact Fermi-type interaction. The latter leads to a  singular potential $W_L \propto 1/r^3$ and logarithmic divergency of the matrix elements for the interaction between electron and quark at $r \to 0$. To allow for a more accurate numerical calculations we present this potential for the finite size $R$ of the nucleus  and cut-off for large momenta (small distances $r$) produced by the  $Z$- boson propagator ($1/(q^2 +M_Z^2$) instead of $1/M_Z^2$  ).    
Full PNC operator has the form
\begin{eqnarray}
 &&W(r)=\frac{G}{2\sqrt{2}}\gamma_5\left[-Q_W\rho(r) \right. \label{e:Qw} \\
 &&\left. +\int d^3r' \rho({\bf r'})\frac{2 Z \alpha
   q m^2 c^2}{3\pi^2 \hbar^2}\frac{I(|{\bf r-r'}|)}{ {|\bf r-r'}|}\right]
  \label{e:wr}\\
 && \equiv  W_Q(r)+W_L(r).
\label{e:wrl}
\end{eqnarray}
Here the first line presents contact PNC interaction  $W_Q(r)$ and the second line is the long-range PNC interaction  $W_L(r)$,
$Q_W \approx -0.9884N+0.07096Z$~\cite{SM} is the weak nuclear charge,   $\rho(r)$ is the nuclear density normalised by condition $\int \rho(r) dV=1$, $\alpha$ is the fine structure constant, and $m$ is the  mass of the fermion in the loop. 
In Eq.~(\ref{e:wr})  the factor $q=(1-4\sin^2\theta_W)$ for electron  and  other charged  leptons ($\mu$, $\tau$) of mass $m$. Quarks also contribute to the potential. For the $u,c,t$ quarks we have factor 
$3q=2(1-\frac{8}{3}\sin^2\theta_W)$; for the $d,s,b$ quarks the factor is $3q=(1-\frac{4}{3}\sin^2\theta_W)$. These factors are the products of the electric and weak quark charges.
They also include factor 3 for 3 possible quark colours.  To reproduce proton weak charge  $q_p=(1-4\sin^2\theta_W)=0.07096$
 including radiative corrections, we use value of the Weinberg angle  near $Z$-pole, $\sin^2\theta_W \approx 0.232$ (formally at zero momentum transfer  $\sin^2\theta_W \approx 0.239$) \cite{SM}. 
 Function $I(r)$  in Eq.~(\ref{e:wr}) is given by  
 \begin{eqnarray}
 &&  I(r)=\\
 \nonumber
&&  \int_1^{\infty}\exp(-2x m c r/\hbar )\left(1+\frac{1}{2x^2}\right) \frac{ \sqrt{x^2-1}z^2 dx}{x^2 +z^2},
\label{e:I}
\end{eqnarray}
 where $z=M_Z/(2m)$. 
Note that  this result takes into account that there is no $Z-\gamma$ mixing for zero momentum transfer. Function $I(r)/r$ gives us dependence of interaction between electron and quark on distance $r$ between them. For large $r$ the function  
 $I(r)/r \propto \exp(-2 m c r/\hbar )/r^{5/2}$,
for $\hbar/(M_Z c) \ll r \ll \hbar/(m c)$  we obtain  $I(r)/r \approx  \hbar^2/(4 m^2 c^2 r^3)$ and this behaviour gives logarithmic divergency of the matrix elements integrated with $d^3 r$. Natural cut-off happens on   $r \ll  \hbar/(M_Z c)$,  where $I(r)/r \propto (\ln r)/r$ and has no divergency integrated with $d^3 r$.  The interval $\hbar/(M_Z c) < r  < \hbar/(M_Z c)$ gives the dominating contribution to the matrix element since it is enhanced by the large parameter $\ln{[M_Z/m]}$.

PNC amplitudes are proportional to  the matrix elements  $\langle s_{1/2}| W_Q + W_L |p_{1/2} \rangle$. 
 Let us start from the approximate analytical calculation  of the ratio of the matrix elements of $W_L$ and $W_Q$. Due to singular behaviour of $I(|{\bf r-r'}|)/|{\bf r-r'}|$ at small distance $|{\bf r-r'}|$  between electron and quark inside the nucleus, we can replace $I(r)/r$ by its contact limit, $I(r)/r \to C\delta({\bf r})$, where $C=\int (I(r)/r) d^3r$. After this substitution operators $W_L$ and $W_Q$ are proportional to each other and we obtain the following result
\begin{equation}
\frac{\langle ns_{1/2}|W_L|np_{1/2}\rangle}{\langle ns_{1/2}|W_Q|np_{1/2}\rangle} \approx  \frac{W_L}{W_Q}
\approx  - \frac{2 \alpha Z S}{3 \pi Q_W} \,,
\label{e:LQ}
\end{equation}
where 
\begin{eqnarray}
\nonumber
&&S=\sum_i q_i L_i\\
\nonumber
&&L_i=(1-\frac{1}{2 z_i^2})(1+\frac{1}{ z_i^2})^{1/2}\ln{[z_i+(1+z_i^2)^{1/2}]}\\
\nonumber
 &&-\frac{5}{6} +\frac{1}{2 z_i^2} \approx
\ln{[M_Z/m_i]} -5/6,
\label{e:S}
\end{eqnarray}
where $z_i=M_Z/(2m_i) \gg 1$. Note that corrections to the last equality are very small, $\sim 1/z_i^2$.
The result has a relatively weak logarithmic sensitivity  to  masses $m_i$.  To have correct exponential cut-off of  the potential $W_L(r)$ at large distance we should select quark mass which provides correct minimal hadron energy  for the system containing quark -antiquark pair.  In the case of $u$ and $d$ quarks this is a pair of pions. Therefore, we  select $2 m_u=2 m_d= 2 m_{\pi}$=280 MeV, the minimal mass of hadrons in the loop on the diagram on Fig.~\ref{fig}~b . Similarly, we choose $2 m_s=2 m_K$= 987 MeV 
(Note that in the  calculations of the radiative corrections to the weak charge  Ref. \cite{Marciano}
 used $2 m_u=2 m_d= 2 m_s \approx 200$ MeV).
 For heavy quark masses we use their bare values $m_c$= 1270 MeV, $m_b$=4500 MeV. 
 A calculation of the hadron loop contribution on  Fig.~\ref{fig}~b could, in principle, be refined using dispersive analysis of $e^+e^-$ annihilation data. However, such approach is not free from uncertainties (see e.g. discussion of the running Weinberg angle  in Ref. \cite{Erler}). This   calculation is beyond the scope of the present paper. 
 
Ratio  $Q_W/Z= -0.9884N/Z+0.07096$ is  approximately the same for all heavy atoms. Numerical estimate of expression Eq. (\ref{e:LQ}) gives correction to PNC amplitude about 2 \%. If we consider the electron loop contribution only, we obtain correction to PNC amplitude  0.1\%.
 
 We have tested analytical result  in Eq. (\ref{e:LQ}), obtained in the contact interaction approximation, by the accurate numerical calculations of the ratio of the  matrix elements of $W_Q$ and $W_L$. Our special interest is in deviation of the accurate result from the contact limit  in Eq. (\ref{e:LQ}). Zero approximation has been calculated  using Hatree-Fock-Dirac relativistic electron wave functions. 
We perform the calculation of the core polarization effect using  the random phase approximation (RPA). Correlation corrections have been included  using the correlation potential method~\cite{DzuFlaSilSus87}.

The effect of $W_Q$ is proportional to  the  $\langle s_{1/2}|W_Q |p_{1/2} \rangle$ matrix elements.
Other matrix elements are negligible in the Hartree-Fock approximation and gain significant values only due to the core polarization corrections,
which are due to the $\langle s_{1/2}|W_Q |p_{1/2}\rangle$ weak matrix elements between the core and excited states. Contrary to the contact PNC interaction $W_Q$, the long-range interaction $W_L$ gives direct contribution  to the matrix elements between electron orbitals with angular momentum  higher than $l=0$ and $l=1$. However, the core polarization contribution still strongly dominates in such matrix elements. For example,  in  $\langle 6p_{3/2}|W_L |5 d_{3/2} \rangle$ matrix element in Cs atom the core polarization contribution is 1000 times bigger than the direct contribution. For Ra$^+$ it is 470 times bigger. 
Therefore, the ratio of the $W_L$ and $W_Q$  contributions to  the PNC effects  is very close to the ratio of   $s_{1/2}-p_{1/2}$ weak matrix elements.
Note also, that it is sufficient to calculate the  ratio $\langle ns_{1/2}|W_L|np_{1/2}\rangle/\langle ns_{1/2}|W_Q|np_{1/2}\rangle$ for any
principal quantum number $n$. This is because the values of these matrix elements come from short distances where the wave functions for different $n$ differ by normalisation only. The normalisation cancels out in the ratio. We use lowest valence states in the calculations. The ratio is also the same for atoms and singly charged ions of these atoms.

The results of calculations for atoms and ions  of experimental interest are presented in Table~\ref{t:rpnc} in a form of the ratio of the (\ref{e:Qw}) and (\ref{e:wr}) parts of the PNC operator, $\langle ns| W_L|np\rangle/\langle ns| W_Q|np\rangle$. We consider two cases, A and B. In case A only the electron loop contribution to the long-range PNC potential (\ref{e:wr}) is included. In case B contributions from all leptons ($e\,\,\mu,\,\tau$) and $u,\,d,\,s,\,c,\,b$ quarks (except for $t$) are included. The reason for separating electron contribution comes from the fact that this is the only true long-range contribution. The distances which give significant contribution to the matrix elements are much larger than the nuclear radius.
The ranges of other contributions  are still much bigger than the range of the weak interaction equal to the $Z$ -boson Compton wave length.  However, their  range is smaller than the nuclear radius and numerically the contributions of $\mu,\,\tau, u,\,d,\,s,\,c,\,b$   may be described very accurately by the contact interaction.


\begin{table}
\caption{\label{t:rpnc}Ratios of PNC matrix elements $\langle ns| W_L|np\rangle/\langle ns| W_Q|np\rangle$ for atoms and singly charged ions of these atoms
calculated  in the contact approximation Eq.  (\ref{e:LQ}) and using accurate relativistic  many body theory. Numbers in square brackets mean powers of ten.}
\begin{ruledtabular}
\begin{tabular}{ccccc}
\multicolumn{1}{c}{Atom}&
\multicolumn{1}{c}{A\footnotemark[1]} &
\multicolumn{1}{c}{A$_c$\footnotemark[2]} &
\multicolumn{1}{c}{(A-A$_c)$/A} &
\multicolumn{1}{c}{B\footnotemark[3]} \\
\hline

$^{40}$Ca   &  1.33[-3] & 1.37[-3]  &  -3.17 \% & 2.84[-2]  \\
$^{85}$Rb   &  9.69[-4] & 1.04[-3]  &  -6.93 \% & 2.15[-2]  \\
$^{133}$Cs  &  8.42[-4] & 9.43[-4]  & -11.99 \% & 1.96[-2]  \\
$^{135}$Ba  &  8.46[-4] & 9.49[-4]  & -12.17 \% & 1.97[-2]  \\
$^{149}$Sm  &  8.37[-4] & 9.54[-4]  & -14.02 \% & 1.98[-2]  \\
$^{163}$Dy  &  7.88[-4] & 9.09[-4]  & -15.25 \% & 1.88[-2]  \\
$^{171}$Yb  &  7.96[-4] & 9.26[-4]  & -16.32 \% & 1.92[-2]  \\
$^{199}$Hg  &  7.54[-4] & 8.97[-4]  & -19.01 \% & 1.86[-2]  \\
$^{203}$Tl  &  7.42[-4] & 8.85[-4]  & -19.37 \% & 1.84[-2]  \\
$^{207}$Pb  &  7.31[-4] & 8.74[-4]  & -19.59 \% & 1.81[-2]  \\
$^{209}$Bi  &  7.33[-4] & 8.78[-4]  & -19.87 \% & 1.82[-2]  \\
$^{213}$Fr  &  7.62[-4] & 9.23[-4]  & -21.09 \% & 1.92[-2]  \\
$^{223}$Ra  &  7.16[-4] & 8.69[-4]  & -21.28 \% & 1.80[-2]  \\

\end{tabular}
\footnotetext[1]{Electron loop contribution only.}
\footnotetext[2]{Contact approximation  for A.}
\footnotetext[3]{Sum of the contributions from $e,\,\mu,\, \tau, \, u,\, d,\, s,\, c,\, b$. The  numerical calculation results are very close  to that given by formula (\ref{e:LQ}). }
\end{ruledtabular}
\end{table}

\section{Ratio of PNC effects in different isotopes}

It was suggested in Ref.~\cite{DzuFlaKhr86} to measure the ratio of PNC amplitudes in different isotopes of the same atom.
It was argued that electronic structure factor cancels out in the ratio and interpretation of the measurements does not require very difficult  atomic calculations which have poor accuracy in atoms with more than one electron in open shells. In fact, the cancelation is not exact and corrections due to the change of the nuclear shape were considered in  Refs.~\cite{BDF09,VF19}. These include the change of the nuclear charge radius and neutron skin corrections. 
 Here we consider one more correction to the ratio which comes from the long-range PNC potential.
We have for the ratio of the PNC amplitudes in isotopes 1 and 2
\begin{equation}
\frac{A_{\rm PNC1}}{A_{\rm PNC2}}  = \frac{\langle ns_{1/2}|W|np_{1/2}\rangle_1}{\langle ns_{1/2}|W|np_{1/2}\rangle_2},
\label{e:ar}
\end{equation}
i.e., it is sufficient to study the ratio of the weak matrix elements.
Let us introduce short notations, $\langle ns_{1/2}|W|np_{1/2}\rangle= Q_WK +   K_L=Q_W K (1+ K_{LK}/Q_W)$. Here $K$ is the electronic structure factor for the  first term in (\ref{e:wrl}),  
$ K_L$ is the matrix element of the long-range PNC potential,  $K_{LK}=K_{L}/K$.  Thus, the relative correction to the single isotope matrix element, presented in Eq. (\ref{e:LQ}), here is denoted by  $K_{LK}/Q_W$.  Then the ratio (\ref{e:ar}) becomes
\begin{equation}
\frac{A_{\rm PNC1}}{A_{\rm PNC2}} = \frac{Q_{W1}K_1 + K_{L,1}}{Q_{W2}K_2 + K_{L,2}}=\frac{K_1}{K_2}\frac{Q_{W1} + K_{LK,1}}{Q_{W2} + K_{LK,2}}.
\label{e:kl}
\end{equation}
It is important  that $K_{LK}=K_{L}/K$ practically does not depend on the isotope, 
while $Q_W$ is approximately proportional to the number of neutrons $N$, so dependence of $Q_W$  on the isotope is significant. 
The  relative difference of the PNC amplitudes for different  isotopes may be approximately presented as  
\begin{equation}
\frac{\Delta A}{A} \approx \left(\frac{\Delta A}{A}\right)_0 (1- K_{LK} /Q_W),
\label{e:kld}
\end{equation}
where $A \equiv A_{\rm PNC}$, $\Delta A =A_1-A_2$, index "0" indicates relative difference of the PNC amplitudes without long-range PNC interaction.
 Thus the correction is equal to $- K_{LK} /Q_W$, so it has opposite sign to  the single isotope correction  $K_{LK} /Q_W$ presented in  Eq. (\ref{e:LQ} ) and Table~\ref{t:rpnc}.

\section{Long range nuclear spin dependent PNC potential}

If we swap $Z$ and $\gamma$ on Fig.~\ref{fig}~b, we obtain a long range PNC potential which depends on  nuclear spin.
Sum of the  nuclear- spin-dependent (NSD)  PNC interaction mediated by the $Z$ exchange \cite{NSDZ} and NSD long range PNC potential  may be presented  in the  following form
\begin{eqnarray}
  &&W(r)=\frac{G}{2\sqrt{2}}  \gamma_0({\bf \Sigma \gamma})\left[(1-4\sin^2\theta_W) \rho(r) \right. \label{e:kappa} \\
  &&\left. -\int d^3r' \rho({\bf r'})
  \frac{2 \alpha q m^2 c^2}{3\pi^2 \hbar^2}        
  \frac{I(|{\bf r-r'}|)}{ {|\bf r-r'}|}\right] , \label{e:wrNSD,}
\end{eqnarray}
where  $\Sigma=1.27 \langle \sum_n\sigma_n-\sum_p \sigma_p \rangle$. The result for the ratio of the long-range  contribution to the $Z$-boson contribution differ from NSI PNC by the numerical  factor $-Q_W/[Z (1-4 \sin ^2 \theta_W)]$. This factor is approximately the same for all heavy atoms.  For Cs this factor is 18.5 and using Table~\ref{t:rpnc} we obtain  the electron loop contribution  1.55\%. In the contact interaction limit it is 13\% bigger (for Fr and Ra$^+$ it is 21\% bigger).  Sum of the contributions from $e,\,\mu,\, \tau, \, u,\, d,\, s,\, c,\, b$  loops increases the $Z$-boson contribution to the NSD PNC effects  by 36\%.  Here the difference with the contact limit is small.

Note that we do not consider here NSD PNC interaction produced by the nuclear anapole moment \cite{anapole,anapole1} and combination of the weak charge and hyperfine interaction \cite{NSDhyperfine}. 

\section{ Long range parity non-conserving potential due to exchange by two neutrinos }

Exchange by two (nearly) massless neutrinos gives  long range potential proportional to $1/r^5$. Parity conserving part of this potential has been calculated in Refs.~\cite{FeinbergPR1968,Feinberg1989,HsuPRD1994}. In addition to the diagram on Fig. \ref{fig}~a, the electron neutrino contribution contains diagrams involving $W$ boson. 
Using their approach we have found parity nonconserving part of this $1/r^5$ potential ($\hbar=c=1$):
\begin{equation}
    W_\nu^{PNC}(r) = -  \frac{G^2}{16\pi^3 r^5}  Q_W (2- N_{eff}) \gamma _5 \,,
\label{e:nuPNC}   
\end{equation}
where $N_{eff}$ is the effective number of the particles with the Compton wavelength larger than $r$.  For molecular scale this is the number of neutrinos, $N_{eff}=3$. However, matrix element of this interaction in atoms converges at very small distances where  $\nu, \, e,\,\mu,\, \tau, \, u,\, d,\, s,\, c,\, b$ contribute giving $N_{eff}=14.6$  (for the parity conserving part of the potential  calculation  of   $N_{eff}$ has been done in Ref.~\cite{StadnikPRL2018}).
 Potential Eq. (\ref{e:nuPNC}) 
is very singular at small $r$ and requires a cut-off parameter $r_c$. In the contact limit of this potential we should replace $1/r^5$ by the integral  $\int d^3r/r^5= 2 \pi /r_c^2$.  
 Potential Eq. (\ref{e:nuPNC}) is applicable for distances $r \gg r_c=\hbar/M_Z c$. If we use this $r_c$ as a cut-off parameter, we find that contribution of the potential Eq. (\ref{e:nuPNC}) to the PNC effects in atoms is $\sim \alpha$, i.e. about  1\%. 

 For a more accurate extension of  this potential to small distances  
we present this potential for the finite size $R$ of the nucleus  and cut-off for large momenta (small distances $r$) produced by the  $Z$- boson propagator ($1/(q^2 +M_Z^2$) instead of $1/M_Z^2$  ).    
Full PNC operator has the form ($\hbar=c=1$) 
\begin{eqnarray}
\nonumber
 &&W(r)=-\frac{G}{2\sqrt{2}}Q_W \gamma_5\left[\rho(r) \right. \\
 &&\left. +(2- N_{eff}) \frac{\sqrt{2}G m^4} {3 \pi^3}\int d^3r' \rho({\bf r'})\frac{I_2(|{\bf r-r'}|)}{ {|\bf r-r'}|}\right]
  \label{e:wr2}\\
 && \equiv  W_Q(r)+W_\nu^{PNC}(r).
\label{e:wrl2}
\end{eqnarray}
For zero nuclear size and $\hbar/(M_Z c) \ll r \ll \hbar/(m c)$ Eq. (\ref{e:wr2}) reproduces Eq.  (\ref{e:nuPNC}) if  
\begin{eqnarray}
 &&  I_2(r)=\\
 \nonumber
&&  \int_1^{\infty}\exp(-2x m c r/\hbar )\left(1+\frac{1}{2x^2}\right) \frac{ \sqrt{x^2-1} x^2 z^4 dx}{(x^2 +z^2)^2},
\label{e:I}
\end{eqnarray}
 where $z=M_Z/(2m)$. 
 Function $I_2(r)/r$ gives us dependence of interaction between electron and quark on distance $r$ between them. For large $r$ the function  
 $I(r)/r \propto \exp(-2 m c r/\hbar )/r^{5/2}$,
for $\hbar/(M_Z c) \ll r \ll \hbar/(m c)$  we obtain  $I_2(r)/r \propto 1/r^5$ and this behaviour gives divergency $1/r_c^2$ of the matrix elements integrated with $d^3 r$, where $r_c$ is the cut-off parameter. Natural cut-off happens on   $r \ll  r_c=\hbar/(M_Z c)$,  where $I(r)/r \propto (\ln r)/r$ and has no divergency integrated with $d^3 r$.  Note that behaviour of the neutrino exchange potential at small distance 
has been investigated in Ref.~\cite{2nushort}. However, they do not study this potential in the standard model. They replaced $Z$ boson by some new scalar particle  and study parity conserving potential only.

Convergence of the integral  in the matrix elements $\langle s_{1/2}|  W_\nu^{PNC}|p_{1/2}\rangle$ on the distance $r \sim r_c=\hbar/M_Z c$ indicates that this interaction in atoms may be treated as a  contact interaction, 
 Due to singular behaviour of $W_L$ at small distance $|{\bf r-r'}|$  between electron and quark inside the nucleus, we can replace $I_2(r)/r$ by its contact limit, $I_2(r)/r \to C\delta({\bf r})$, where $C=\int (I_2(r)/r) d^3r$. 
After calculation of the contact limit of $I_2(r)/r$ we obtain  potential    $W_\nu^{PNC}(r)$ which  is proportional to the weak interaction mediated by $Z$-boson in Eq.  (\ref{e:Qw}). Therefore, we may present the result for the relative correction to the PNC amplitude as 
\begin{equation}
 \frac{   W_\nu^{PNC}(r)}{W_Q(r)} =  - \frac{G  M_Z^2}{12\sqrt{2}\pi^2} (N_{eff}-2)=-0.72 \%\,,
\label{e:nuPNCvRatio}   
\end{equation}
 This estimate of $W_\nu^{PNC}(r)$  contribution significantly exceeds the experimental error 0.35 \% for Cs PNC amplitude.
However, we may assume that a greater part of this correction has already been included among radiative corrections to the weak charge $Q_W$.

In principle, one may think about some macroscopic effects produced by the PNC potential Eq.~(\ref{e:nuPNC}). Such experiments have been done for the parity conserving  potentials - see  Refs.~\cite{Kapner2007,Adelberger2007,Chen2016,Vasilakis2009,Terrano2015}. However, rapid decay with the distance indicates that corresponding effects will be very small.

\section{Conclusion}

We calculated the long range PNC potentials described by the diagram Fig.~\ref{fig}~a ($\propto 1/r^5$) and Fig.~\ref{fig}~b  ($\propto 1/r^3$). These potentials   contribute to  the PNC effects in atoms and molecules. Contrary to the contact weak interaction, these potentials may mix opposite parity orbitals with orbital angular momentum higher than $l=0$ and $l=1$, but $s_{1/2}$ -$p_{1/2}$ mixing still gives a dominating contribution.  Contribution of the $1/r^3$ potential on Fig.~\ref{fig}~b to the nuclear spin independent PNC effects is  2\%, the contribution to  the nuclear-spin-dependent effects is 40\% of the $Z$-boson contribution.  However, similar Feynman diagrams have already been included as the radiative corrections to the weak charge $Q_W$ which is the source of the contact PNC interaction in atoms and molecules. Therefore, we may assume that only deviation from the contact approximation is an additional contribution to PNC effects. Diagram on Fig.~\ref{fig}~b  with electron loop gives the PNC interaction range which exceeds the weak interaction range due to  Z-boson exchange $M_Z/(2 m_e)=10^5$ times. However, the electron loop contribution is only 0.1\% of the weak charge $Q_W$ contribution. For nuclear-spin-dependent PNC interaction the electron loop contribution is 2\% of the Z-boson contribution. Contributions of other charged fermions to the PNC matrix elements are very close to the contact limit since the range of corresponding interactions is smaller than the nuclear size.  

Integrals in the matrix elements  of the $1/r^5$ potentials are  dominated by very small $r$ and corresponding interaction is accurately presented by its contact limit. Therefore, its effects may be treated as the radiative corrections to the weak charge $Q_W$ and $\kappa_2$, which are the strength constants of the contact nuclear spin independent and nuclear spin dependent weak interaction. 

In the paper Ref.~\cite{StadnikPRL2018} the parity conserving part of the potential  $1/r^5$ have been considered and compared with experimental data on muonium, positronium, hydrogen and deuterium  spectra and isotope shifts in hydrogen and calcium isotopes.   The results have been expressed as limits on the interaction constant denoted as  $G_{eff}$. These limits are several orders of magnitudes weaker  than the calculated interaction constant within the Standard model (including $\nu, \, e,\,\mu,\, \tau, \, u,\, d,\, s,\, c,\, b$ particles in the loop on the diagram  on Fig.~\ref{fig}~a), from  $G_{eff}^2/G^2 < 4.0 \cdot 10^{11}$ to $G_{eff}^2/G^2 <  1.9 \cdot 10^{2}$. The latter limit is 18 orders of magnitude better than the limits obtained from macroscopic experiments  Refs.~\cite{StadnikPRL2018,Kapner2007,Adelberger2007,Chen2016,Vasilakis2009,Terrano2015}.
 
  The situation with the parity non-conserving parts of the long-range potentials considered in the present work is more optimistic. If following Ref.~\cite{StadnikPRL2018}  we treat the interaction constant as a phenomenological parameter characterising some interaction beyond the standard model,   then from the Cs PNC experiment we obtain $G_{eff}^2 <  0.3 G^2$ for the $1/r^3$ potential and $G_{eff}^2  <  G^2$ for the $1/r^5$ potential (theoretical and experimental errors have been added in quadrature). 
   
\section*{Acknowledgements}
 We are grateful to M. Pospelov and E. Shuryak for valuable discussions and to Xunjie Xu for attracting our attention to Refs.  \cite{Ghosh} and \cite{2nushort}.  This work was supported by the Australian Research Council Grants No. DP190100974 and DP200100150 and the Gutenberg Fellowship.



\end{document}